\documentclass[runningheads]{llncs}
\usepackage{xcolor}
\newcommand{\tj}[1]{\textcolor{black}{#1}}

\newcommand{\fj}[1]{\textcolor{black}{#1}}
\newcommand{\iocrc}[1]{\textcolor{black}{#1}}

\newcommand{\gm}[1]{#1}
\newcommand{\bj}[1]{#1}
\newcommand{\cj}[1]{#1}
\newcommand{\ya}[1]{#1}
\newcommand{\io}[1]{#1}

\usepackage{subcaption}
\usepackage{booktabs}
\usepackage{algorithm}
\usepackage{algpseudocode}
\usepackage{amsmath}
\usepackage{graphicx}
\usepackage{diagbox}
\usepackage{booktabs}
\usepackage{vcell}
\usepackage{multicol}
\usepackage{array}
\usepackage{adjustbox}
\usepackage{hyperref}
\usepackage{orcidlink}

\usepackage[noadjust]{marginnote}
\newcommand{\pageenlarge}[1]{\enlargethispage{#1\baselineskip}}

\begin{document}

\title{Query Exposure Prediction for Groups of Documents in Rankings}
\author{Thomas Jaenich, Graham McDonald, Iadh Ounis}
\titlerunning{Query Exposure Prediction for Groups of Documents in Rankings}
\institute{University of Glasgow, Glasgow, UK\\
\email{t.jaenich.1@research.gla.ac.uk}\\
\email{\{graham.mcdonald,iadh.ounis\}@glasgow.ac.uk}
}

\maketitle              
\begin{abstract}
\looseness -1 The main objective of an \ya{Information Retrieval (IR)} system is to provide a user with the most relevant documents to the user’s query. To do this, modern IR systems typically deploy a re-ranking pipeline in which a set of documents is retrieved by a lightweight first-stage retrieval \iocrc{process} and then re-ranked by a more effective but expensive model. However, the success of a re-ranking pipeline is heavily dependent on the performance of the first stage retrieval, since new documents are not usually identified during the re-ranking stage. Moreover, this can impact the amount of exposure that a particular group of documents, such as documents from a particular demographic group, can receive in the final ranking. 
 \ya{For example, the fair allocation of exposure becomes more challenging or impossible if the first stage retrieval \iocrc{returns} too few documents from certain groups, since the number of group documents in the ranking affects the exposure more than the documents' positions.} With this in mind, it \gm{is} \ya{beneficial} to predict the amount of exposure that a group of documents is likely to receive in the results of the first stage retrieval \iocrc{process}, \iocrc{in order} to ensure that there are a sufficient number of documents included from each of the groups. In this \ya{paper}, we introduce the novel task of query exposure prediction (QEP). Specifically, we \ya{propose the first approach for} predicting the distribution of exposure that groups of documents will receive for a given query. 
 \ya{Our \iocrc{new} approach, called GEP,} uses lexical information from individual groups of documents to estimate the exposure the groups will receive in a ranking. Our experiments on the TREC 2021 and 2022 Fair Ranking Track test collections show that our proposed GEP approach results in exposure predictions that are up to $\sim$40\% more accurate than the predictions of suitably adapted \iocrc{existing} \ya{query performance \gm{prediction} (QPP)} and resource allocation approaches.

\end{abstract}

\section{Introduction}\label{sec:intro}
\pageenlarge{1}
\tj{\ya{The} main objective of \ya{an information retrieval (IR)} system \ya{is} to provide \ya{the} users with the documents that are the most relevant to \ya{their} search query. \ya{Recently}, deep neural ranking models that use \ya{contextualised} language representations, such as BERT~\cite{devlin2018bert}, have been \iocrc{shown to effectively score} documents and determine their relevance to a user's query~\cite{lin2022pretrained}. However, while effective, applying these neural models is \gm{computationally} expensive and \gm{is} often infeasible over large collections~\cite {hofstatter2019let}. Therefore, deep neural ranking models are usually deployed in re-ranking pipelines. In a re-ranking pipeline, \bj{an inexpensive} ranking model, such \ya{as} BM25~\cite{robertson1995okapi} or the more recent SPLADE~\cite{formal2021splade}, is used to create an initial pool of documents in a first-pass retrieval \iocrc{process}. The documents from this pool are \ya{then} re-scored by deep neural ranking models such as ELECTRA~\cite{pradeep2022squeezing} or T5~\cite{pradeep2021expando}. \gm{The re-scored documents are then} presented to the user \ya{in decreasing order of relevance}.
However, the amount of exposure to a user that a document is likely to receive is dependent on the document's position in the ranking, since according to the well-known position bias model~\cite{craswell2008experimental}, lower-ranked documents have a lower probability of being exposed to, or examined by, a user. This can lead to an unintended bias against particular \textit{groups} of documents. For example, if relevant documents that are associated with a demographic group, such as race or gender, are systematically ranked at lower positions, then the group will receive less exposure to the user than the groups that have documents ranked at higher positions~\cite{singh2018fairness}. In re-ranking pipelines, the exposure distribution in the final ranking is \ya{typically} limited by the recall of the initial candidate pool, i.e., documents \gm{that are} not retrieved by \iocrc{the} first-pass retrieval \iocrc{process} \gm{are not} scored by the re-ranker. If the initial pool of documents contains only \ya{a} few or \ya{no} relevant documents from a group, \ya{a} fair distribution of exposure in the final ranking might become infeasible.} \fj{For example, Bower et al.~\cite{bower2022random} have shown that inequality in the initial pool of documents in a  multi-stage ranking process can hinder the effectiveness of existing fairness mitigation methods, such as \iocrc{randomisation}~\cite{diaz2020evaluating}.}
\ya{Therefore, predicting the exposure of a group before deploying the initial ranker \iocrc{is both important and} useful. Indeed, this would, \iocrc{for example}, allow \iocrc{the system} to adjust the weights of different groups in the first-pass retrieval model, so \iocrc{as to} have enough documents from each group in the candidate pool, and a fair chance of achieving the desired exposure in the search results when a re-ranker \iocrc{is applied}}.

\pageenlarge{1}
\looseness -1In this work, we introduce the \ya{new} task of Query Exposure Prediction (QEP), i.e., predicting how the available exposure in search results will be distributed among groups of (related) documents in response to a user's query. Developing exposure distribution prediction models will enable system designers to choose an appropriate initial ranker and deploy fairness strategies on a case-by-case basis, i.e., when the predicted exposure distribution is not closely correlated with a desired, or target, distribution. \ya{Specifically, we} propose the first QEP approach, \ya{called} \bj{Group Exposure Predictor (GEP)}. Our \iocrc{new} approach is inspired by pre-retrieval Query Performance Prediction (QPP)~\cite{he2006query} approaches that aim to estimate how difficult it will be for an IR system to provide relevant documents in response to a particular query. However, different\gm{ly} from pre-retrieval QPP \ya{approaches}, our proposed GEP approach leverages statistics about how a query's terms are distributed in the indexed documents from each group to estimate how the available exposure will be distributed among the different groups in a ranking.  \ya{We evaluate the performance of GEP in two scenarios. One where the query is passed to the predictor and executed as it is, and another where the query is expanded by Pseudo-Relevance Feedback (PRF) methods before being passed to the predictor.}
\ya{Our extensive experiments on} the TREC 2021 \& 2022 Fair Ranking Track~\cite{trec-fair-ranking-2021,trec-fair-ranking-2022} \gm{test collections} show that our \iocrc{proposed} exposure prediction model GEP is able to predict the distribution of exposure better than traditional performance predictors as well as an established \ya{federated} information retrieval approach. \bj{Specifically, our predictor shows statistically significant improvements over the baselines for the majority of observed exposure distributions, with decreases in the prediction error of up to $\sim$40\% compared to the next best \ya{baseline}. }



\section{Related Work}\label{sec:related}
\pageenlarge{1}
\ya{A fair allocation of document exposure in a ranking is increasingly recognised as a key objective besides relevance when developing search systems}~\cite{ekstrand2019fairness}. Specifically, in the fairness literature, many \ya{prior works}~\cite{biega2018equity,heuss2022fairness,jaenich2023colbert,kletti2022pareto,morik2020controlling,sarvi2022understanding,singh2018fairness,usunier2022fast,zehlike2020reducing} \ya{have} proposed approaches to avoid unfair distributions of exposure in rankings.
Such approaches have typically focused on fairly distributing the exposure over individual documents or over predefined groups of documents~\cite{zehlike2021fairness}. In this work, we are concerned with groups of documents in a ranking. The definition of the groups of documents in a ranking usually depends on the search task at hand. In many cases, the documents are grouped by an associated protected characteristic. The protected characteristic, sometimes also referred to as a protected attribute, is defined as an associated label to \ya{a given} document, \ya{which} \gm{should not} be used as the basis of decision-making. \ya{Some prominent} examples of protected attributes \ya{include} geographic locations, ethnicity, \ya{and} gender.
\gm{Among} the many studies that propose mitigation approaches, \ya{several} works have \ya{suggested} how to measure fairness in terms of exposure~\cite{diaz2020evaluating,raj2020comparing,wu2022joint}. \ya{However, we are not aware of any previous work on predicting the exposure of groups in a ranking. }
Therefore, we \ya{aim} to fill this research gap by \iocrc{proposing} the first query performance predictor for \ya{measuring the exposure of} groups of documents.
Query Performance Prediction (QPP) is traditionally applied to \ya{predict the effectiveness of the ranking produced by an} IR system \ya{in response to a query} without using human-relevance judgements. This is useful since the performance of IR systems varies over different queries. \ya{The} QPP \ya{predictors} are usually separated in \textit{post-retrieval} and \textit{pre-retrieval} \ya{methods}. Post-retrieval methods are applied after an initial ranking was created \ya{by the system} and therefore have access to \ya{the} estimated relevance scores, \ya{which} can be used for \ya{the performance} prediction~\cite{carmel2010estimating}. However, post-retrieval \ya{methods} are not applicable \ya{in our scenario}, since after \ya{the} initial retrieval stage, \ya{the} exposure can be directly calculated, rendering \ya{the} prediction redundant. 
\ya{On the other hand,} pre\gm{-}retrieval query performance prediction \ya{methods} traditionally use properties from the underlying document corpus or information obtained from the query terms \ya{to make} their predictions. 
\ya{For example,} average inverse document frequency, AvIDF~\cite{cronen2002predicting}, \ya{approximates} the average specificity of query terms \ya{by} calculating the inverse document frequency (idf) \ya{for each} query term. Related to this, He et al.~\cite{he2004inferring} have introduced the average inverse collection term frequency (AvICTF). AvICTF is similar to AvIDF but uses the term frequency \bj{in a collection} instead of the document frequency in the calculation of IDF. Further extending the concept of AvICTF, He et al. \ya{proposed} to use the Simplified Clarity Score (SCS). \cj{\bj{SCS measures how ambiguous a query is. SCS is calculated using the sum of the AvICTF and a maximum likelihood estimate on how often a term appears in a query.}} These predictors \ya{serve as} baselines \ya{in} our experiments\gm{.}
\ya{Different}\gm{ly} from predictors \ya{that solely use} lexical information of the \bj{query terms from the collection statistics}, the term-relatedness predictor, Averaged Pointwise Mutual Information (AvPMI)~\cite{hauff2010query}, calculates the probability of two terms appearing together in a document. \cj{AvPMI has been shown to be a robust predictor due to its ability to exploit co-occurrence statistics \iocrc{in} a collection. \iocrc{Therefore,} we \gm{include} it as a baseline in our experiment\gm{s},} \gm{and} \ya{evaluate how well the \bj{query} performance predictors \gm{AvIDF, AvICTF, SCS and AvPMI} can estimate the exposure of groups of documents in a ranking.}

In a related manner, resource selection approaches such as the prominent CORI~\cite{callan2000distributed} algorithm from federated search,  predict the suitability of different data sources for \iocrc{retrieving content for a given query}. Conceptually similar \ya{to the task we are tackling}, the CORI algorithm uses a probability denoted as \cj{\textit{belief}}, which is calculated from \ya{the} collection statistics \gm{and} \ya{predicts} how suitable a data source is \gm{for} \ya{a given query}. \gm{W}\ya{e} adapt this algorithm \ya{to our task and use it as one of our} baselines. \fj{In \iocrc{this} work, we \iocrc{particularly} focus on and adapt \iocrc{QPP methods} based on lexical information. Unlike neural approaches, \iocrc{lexical QPP} predictors rely on clear and interpretable features derived from lexical information. This makes it easier to \iocrc{capture} the underlying mechanisms contributing to the prediction as well as developing and evaluating a new predictor \iocrc{for our new query exposure prediction task}.} \tj{\ya{Next, before the introduction of} our \ya{proposed} predictor, we \ya{first describe in more detail} the exposure distribution problem.}

\section{Exposure in Rankings}

\begin{figure*}
    \centering
    \includegraphics[width=0.72\textwidth]{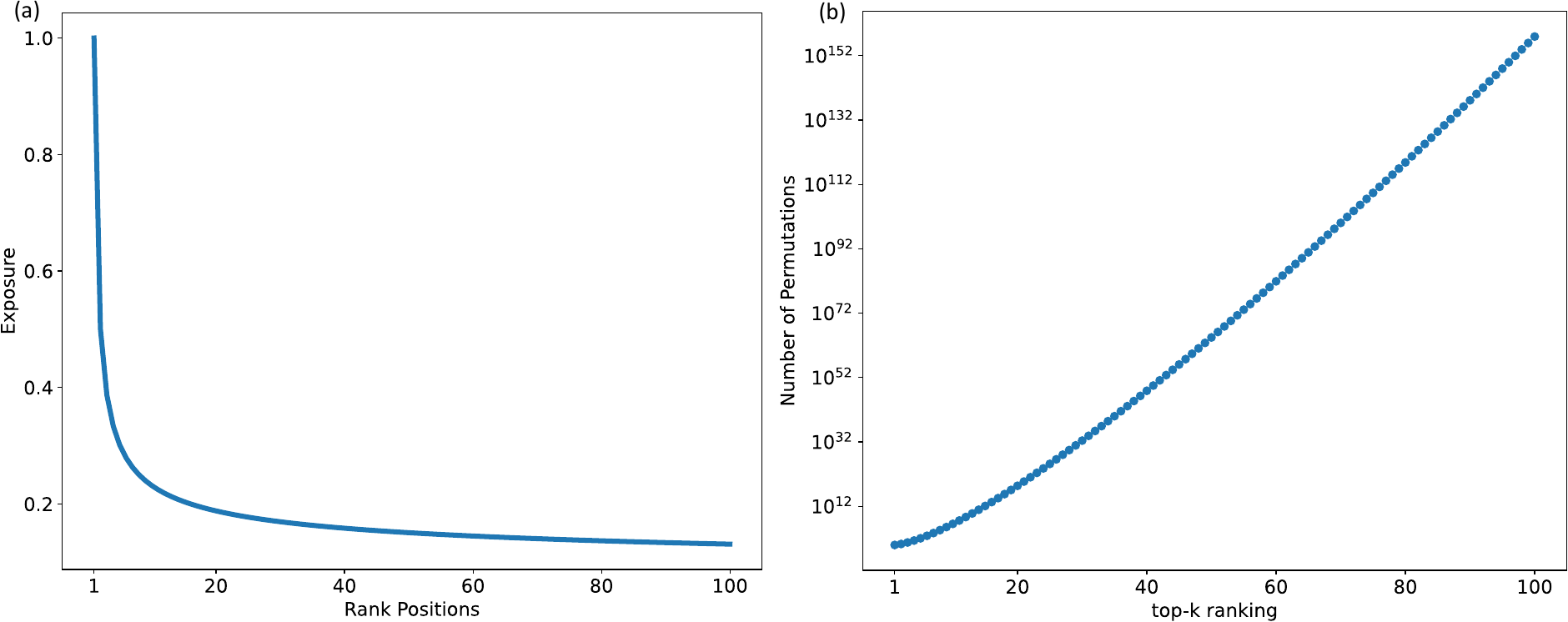}
    \caption{Plot (a) shows the amount of available exposure for each position in a ranking of size $k\text{=}100$ according to the position bias model from the well-known DCG metric~\cite{jarvelin2002cumulated}. Plot (b) shows the number of possible orderings of documents for rankings of size $k\text{=}{1...100}$. The y-axis of Plot (b) is \bj{in} log scale.}
    \label{fig:exposureCombined}
\end{figure*}

In this work, \ya{we aim to predict how much exposure \gm{to a user} a group of documents will receive when the documents are ranked by \gm{their} relevance to the user's query}.
\ya{The expected exposure of a given document depends on its position in the ranking, since a user usually starts from the top and examines each document in order with a certain probability of stopping based on the number of relevant (or partially relevant) documents they have seen.} Therefore, the further down \ya{in} the ranking \ya{the document is}, the less likely \ya{it will be} exposed to the user (often referred to as \ya{the} position bias~\cite{craswell2008experimental}). There have been a number of \bj{user models}, \ya{which have been} proposed in the literature that estimate the likelihood of a user examining a \ya{document at a} particular rank position, for example~\cite{agarwal2019estimating,wang2018position,craswell2008experimental}. In this work, \ya{we estimate the} available \ya{document} exposure \ya{at a given} position in the ranking \ya{using} the standard exposure drop-off (position bias) from the well-known Discounted Cumulative Gain (DCG)\cite{jarvelin2002cumulated} metric, defined as:  $\frac{1}{(log_2(p)+1)}$
where $p$ is a position in the ranking. Figure~\ref{fig:exposureCombined}(a) illustrates the exposure distribution that is available for a ranking of length $k\text{=}100$ under \ya{the} DCG position bias/exposure model. Figure~\ref{fig:exposureCombined}(a) \ya{shows that} the \ya{available} exposure \ya{drops sharply} in rank positions 1..10 \ya{and then} decreases \ya{more gradually} in positions 11..100. Figure~\ref{fig:exposureCombined}(b) shows the number of possible orderings for rankings of size $k\text{=}1..100$. As can be seen from Figure~\ref{fig:exposureCombined}(b), there are 10\textsuperscript{152} different possible orderings of documents in a ranking of size $k\text{=}100$. 
\vspace{-1\baselineskip}

\begin{figure*}[tb]
    \centering
    \includegraphics[width=\textwidth]{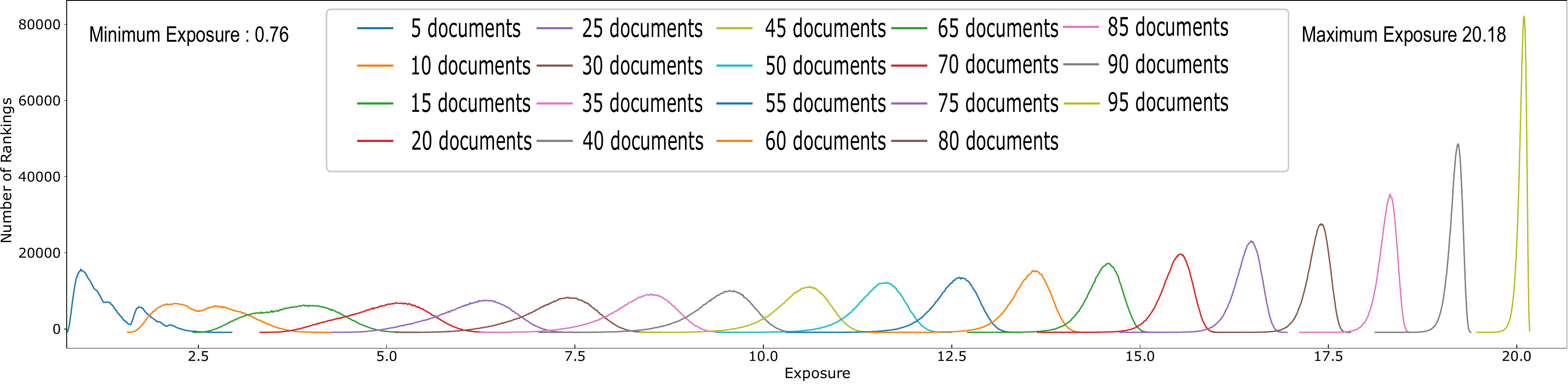}
    \caption{The number of rankings that can achieve a particular amount of exposure for a given number of documents in a single group for a ranking of size $k$=100.}
    \label{fig:possibleExpPerm}
\end{figure*}

The amount of exposure that a group of documents receives is \ya{calculated by} \gm{summing} the exposure \ya{values} for each position in the ranking \ya{where} a document from the group \ya{appears}. \tj{For example, if a group has two documents in a ranking, with the first document \ya{placed} at position 1 (i.e., the top rank position) and the second document placed in the 5\textsuperscript{th} position, the first document receives an exposure of 1 and the second document receives an exposure of 0.39. \ya{In such a scenario, the group will receive an exposure of} $1+0.39=1.39$. \gm{Such an} exposure model \ya{only} considers \ya{the rank of a document to determine its} amount of exposure. \ya{In this case}, the amount of exposure that a group of documents will receive is dependent on (1) the size of the ranking, (2) the number of documents from the group that are in the ranking, and (3) the positions in the ranking \ya{where} the documents from the group \ya{appear}. However, \bj{there are multiple possible rankings (or orderings) that produce the same amount of exposure for a group.} Returning to our previous example of two documents from a group in rank positions 1 and 5, if the \iocrc{positions} of the two documents in the ranking \iocrc{are} swapped then the group's exposure is still $1.39$. }


\looseness -1 \ya{For a ranking of size $k\text{=}100$}, Figure~\ref{fig:possibleExpPerm} presents (1) the range of exposure values that are achievable for a single group depending on the number of documents from the group that are included in the ranking, and (2) the number of possible rankings that achieve that particular amount of exposure. From Figure~\ref{fig:possibleExpPerm}, we can see that the range of achievable exposure is larger when there \ya{is} a relatively small number of documents from a group in the ranking, but \ya{that} the amount of achievable exposure increases as more documents from the group are added to the ranking. Moreover, \ya{the number of documents in the ranking affects the} specific range of achievable exposure. \ya{Furthermore,} certain exposure values are more likely to be achieved given the number of documents in the ranking. 
\ya{Hence, adding more documents from a group to the ranking can notably increase the group's exposure, much more than just moving the existing documents to higher ranks}. \tj{This highlights the importance of the first-pass retrieval and the impact of \ya{low recall on the} re-ranking pipeline 
\ya{since the exposure that a group receives will be much more enhanced by adding more documents from a group in the first-pass retrieval than re-ranking the documents in the existing ranking}. Moreover, this suggests that \ya{enhancing recall using techniques such as} pseudo-relevance feedback (PRF)~\cite{amati2002probabilistic,amati2003probabilistic,abdul2004umass,fang2006semantic} \ya{has} the potential to notably impact the distribution of exposure.}

\section{Group Exposure Prediction - GEP }
\pageenlarge{1}
\bj{Our proposed GEP predictor estimates how the available exposure will be distributed over groups, $g \in G$, of documents\gm{,} in a ranking $r$ \gm{of size $k$} in response to a \gm{user's} query $Q$ with query terms $t_0..t_n$. To predict how the available exposure will be distributed over the groups, for each group, $g \in G$, we first create a vector representation $S_g$, where $|S_g|=n$. The vector $S_g$ contains one score\gm{,} $s$\gm{,} per query term\gm{,} $t \in Q$\gm{, where $s_i$ is the mean of the $k$ largest \textit{tf}-\textsf{idf} scores for the documents in $g$ and the query term $t_i$}.} 
\cj{To calculate $s$ we first calculate the \textit{tf}-\textsf{idf} scores for every document within a group $d_{g,i}$ for a term $t_i$\gm{, where} the inverse document frequency is calculated as:  $\text{idf}(d_g) = log(\frac{||d_g||-df_g+0.5}{df_g+0.5})$\gm{, and} $||d_g||$ is the number of documents that exist for a group $g$ in the collection\gm{,} $df_g$ is the document frequency, i.e., the number of \gm{group} documents the term appears in. In the next step, we collect the term frequencies $tf$, i.e., the number of term occurrences in each document $d_{g,i}$ and combine it with the \textsf{idf} score \textit{tf}-\textsf{idf} = $tf \cdot idf$. To obtain the score $s$  from the calculated \textit{tf}-\textsf{idf} values, we use only the $k$ largest \textit{tf}-\textsf{idf} scores, where $k$ refers to the length of the ranking $r$ for which we predict the exposure distribution. In other words, the maximum number of group documents that could be included in the ranking. If a term appears in less than $k$ documents, we consider the missing scores to be zero. \cj{We only consider the $k$ highest scores because the ranking length is essential for the exposure the documents receive.} \cj{For example, let \ya{us} assume that for a considered group, only one document contains the query term $t_i$. We obtain only one \textit{tf}-\textsf{idf} score, which is potentially very high. However, a group can only occupy a single high position in the final ranking with a single document, so the overall exposure is still low.} Therefore, to avoid any under or over-estimation of $s$ we use the average of the top-$k$ scores, calculating  $i = \frac{1}{k}\sum_{i=1}^{k} \textit{tf-}\textsf{idf}$. For every query term, we store $s$ in the vector $S$. Conceptually, $S$ is a suitable representation of the group indicating which query terms are most important. To use $S$ to make a performance prediction for the group, we need to check its similarity to the original query $Q$. We use the traditional way of calculating \textit{tf}-\textsf{idf} for each $t$ in $Q$. However, instead of calculating the scores only over documents from one group we calculate them over the whole collection. This is necessary since the query will be executed over all of the documents. To compute the similarity $e_g$ for each group to the query we calculate the dot product between $S$ and the query $Q$ : $ e_g = \mathbf{I_g} \cdot \mathbf{Q}$. A higher similarity indicates a higher predicted exposure. However, since the exposure is relative to the other groups we need to calculate the similarity for every group \gm{$g \in G$}. We calculate our final prediction score $GEP_g$ by comparing the individual scores $e_g$ and \ya{normalise} them to be between 0 and 1. This results in a predictor score per group that can be used to estimate the exposure a group will \ya{obtain}. }

\section{Experimental Setup}
\pageenlarge{1}
To evaluate our proposed GEP predictor \ya{in} the task of query exposure prediction, we formulate the following two research questions: 
\begin{enumerate}
    \item  \textbf{RQ1:} Is our proposed GEP approach effective for predicting the exposure that groups will receive when an \bj{inexpensive} ranking model is deployed?
    \item  \textbf{RQ2:} Is GEP effective for predicting the exposure that groups receive in a ranking, when \iocrc{the} query is expanded using pseudo-relevance feedback?
\end{enumerate}
To conduct our experiments and answer our research questions, we use the PyTerrier~\cite{macdonald2020declarative} Information Retrieval Framework.

\noindent\textbf{Test collections:} In our experiments, we use the TREC 2021 and 2022 Fair Ranking Track test collections~\cite{trec-fair-ranking-2021,trec-fair-ranking-2022}. Both test collections consist of English language Wikipedia articles. \iocrc{These test} collections were created to encourage research on how editors of Wikipedia can be presented with a fair ranking of articles that need editing\gm{,} \ya{in relation to} a topic of their interest. In the TREC Fair Ranking Track, fairness \ya{is} measured in terms of exposure. \ya{Hence}, the \iocrc{corresponding test} collections are well suited for investigating our research questions. The \iocrc{test} collections consist of a set of queries as well as fairness group category labels for the documents in the \iocrc{underlying} collection. \bj{The labels categorise \iocrc{the} different features \iocrc{of documents} such as \iocrc{their} popularity or creation date. Both TREC collections were indexed using a Porter Stemmer and stop-word removal. In the next paragraph, we explain the \gm{categories}.}

\looseness -1 \bj{\noindent \textbf{Categories:} \gm{Figure}~\ref{fig:tablemerge} provides an overview of the available categories and their corresponding groups \ya{for which} we aim to predict the exposure. In the TREC 2021 Fair Ranking Track collection, \gm{only the geographic location category has labels for all of the documents in the collection}. Therefore, we use the geographic category in our experiments. \cj{For the TREC 2022 collection, we use the categories proposed by the organisers with group labels for every document in the collection.} The first category is \textit{Age of an article}, which denotes the length of time the article \iocrc{(or document)} has existed. \ya{The age} is mapped into four discrete categories shown in \gm{Figure}~\ref{fig:tablemerge}. \gm{W}e \gm{also} use the categories: \textit{Age of Topic},  \textit{Alphabetical} and \textit{Popularity}. Age of Topic denotes the age of the \gm{main} topic \gm{that is} discussed in an article. Alphabetical categorises articles by the initial letter of \gm{an article's} title. The \textit{Popularity} category labels articles based on their viewing frequency in February 2022.}
\begin{table}[]
    \centering
      \begin{tabular}{l}
     \toprule
         \textbf{Geo-Location} \\
         Unknown, Europe, Africa,  Northern America, 
         Asia, Oceania, \\ Latin America and the Caribbean, Antarctica \\
         \midrule
        \textbf{Age of Article} \\
        2001-2006, 2007-2011, 2012-2016, 2017-2022  \\
        \midrule
       \textbf{Popularity} \\
        Low, Medium-Low, Medium-High, High  \\
        \midrule
        \textbf{Alphabetical} \\
        a-d, e-k, l-r, and s-z  \\
        \midrule
        \textbf{Age of the topic} \\
        Unk, Pre-1900s, 20th century, 21st century \\
    \bottomrule
    \end{tabular}
    \caption{Categories and their groups.}
     \label{fig:tablemerge}\vspace{-2\baselineskip}
\end{table}

\noindent\textbf{Queries:} We use all of the evaluation queries from the two Fair Ranking \gm{Track} collections. Specifically, we use the 49 evaluation queries in the TREC 2021 Fair Ranking \gm{Track} collection and the 47 evaluation queries from \gm{the} 2022 \gm{collection}. All queries consist of keywords relating to a Wikipedia topic, e.g., a query can be a collection of words such as ``museum baroque librarian architecture library".

\noindent\textbf{Retrieval models:} In our experiments, we use a selection of prominent first-pass ranking models commonly used in re-ranking pipelines. We evaluate our predictor on  BM25~\cite{robertson1995okapi} (with default parameters), TF-IDF~\cite{salton1988term} and SPLADE~\cite{formal2021splade}.

\noindent\textbf{PRF:} To evaluate the impact of pseudo-relevance feedback \bj{on the prediction performance of GEP}, we \ya{use} \ya{the} RM3 relevance language model~\cite{abdul2004umass}, 
and KLQueryExpansion~\cite{amati2003probabilistic} (KLQ) in our experiments, with \ya{the} number of feedback documents \ya{set to} 3 and  \ya{the} number of expansion terms \ya{set to} 10.

\noindent\textbf{Baselines:} As baselines, we \ya{use} the CORI~\cite{callan2000distributed} federated \ya{IR algorithm} as well as \ya{several} pre-retrieval Query Performance Predictors. We use Average Inverse Collection Term Frequency (AvICTF)~\cite{he2006query}, Average Inverse Document Frequency (AvIDF)~\cite{cronen2002predicting}, Averaged Pointwise Mutual Information (AvPMI)~\cite{hauff2010query} and Simplified Clarity Score (SCS)~\cite{he2004inferring} as baselines. \cj{ To have a fair comparison with our \iocrc{proposed} approach, we create a prediction score for each group and then \ya{normalise} these group scores to be between 0 and 1.} The normalised scores are used as \ya{a way to measure how much} exposure a group will \ya{receive} in the ranking. \gm{We assume that if a query is predicted to perform better on documents of one group,
compared to the documents of another group, then the first group is likely to receive more exposure in the ranking.} We normalise the prediction scores, since the predictions between groups are dependent i.e., if all prediction scores between groups are high, then the difference in exposure will likely be low. If the predicted performance of a query is high over one group of documents but low on \gm{the} others, then it is likely that the high-performing group will receive most of the exposure in the ranking. \bj{As introduced in the related work section, the CORI algorithm contains a belief component, which depicts how likely a group of documents contains relevant information to a query. In \iocrc{this} work, we choose to use an average of the belief values and a value of 0.4 for the belief parameter in the algorithm, as suggested in the original paper~\cite{callan2000distributed}}.

\noindent\textbf{Measures:} To evaluate our predictions, we calculate the actual exposure the groups \ya{receive} in a ranking and compare \bj{the exposure} to our predicted values. Following the TREC Fair Ranking Track, we use the Jensen-Shannon Distance for \ya{the calculation}. Jensen-Shannon Distance calculates the distance between two probability distributions. We use it to calculate the distance between the distributions of our predicted exposure $P$ and the actual Exposure $E$\iocrc{:} 
\begin{equation}
     JSD(P \| E) = \sqrt{\frac{1}{2} \text{KL}(P \| M) + \frac{1}{2} \text{KL}(E \| M)}
\end{equation}
\cj{where $JSD(P \| E)$ is the distance between the predicted exposure $P$ and the actual exposure $E$. $KL(P \| M)$ is the  Kullback-Leibler (KL) divergence between $P$ and the average distribution $M$. $KL(E \| M)$ is the  Kullback-Leibler divergence between $E$ and $M$. $M$ denotes the midpoint distribution between $P$ and $E$. $M = \frac{1}{2} (P + E)$. If there are no differences between the actual exposure and the predicted value, the distance is 0.}
In our experiments, we consider a predictor to be better when it produces values closer to 0 compared to another predictor. \fj{In practice, when choosing a predictor for deployment, \iocrc{we usually set a target threshold} to \iocrc{ascertain} its usefulness. For example, a predictor \iocrc{should show} a high degree of similarity (e.g., \iocrc{$<$0.2}) with the \iocrc{real} values of a test set before being deployed. However, the \iocrc{choice of} threshold depends on the \iocrc{predictor's application} and the \iocrc{fairness definition in that} context. \iocrc{Hence, we leave such a study to future work}. When analysing our experiments, we test the results of GEP for statistically significant differences to the baselines using a paired t-test (p$<$0.01) with Bonferroni correction.}


\section{Results}
\pageenlarge{1}
In this section, we answer both our research questions \ya{and draw further} insights into the problem of exposure prediction. 

\begin{table*}[tb]
\caption{Evaluation of the predictors on common first-pass retrieval rankers, for rankings with $k$=100. * indicates significant differences (t-test, with Bonferroni correction, p$<$0.01) in prediction performance \iocrc{between} GEP \iocrc{and} the baselines. }
    \centering
      \begin{adjustbox}{width=0.99\textwidth}
  \begin{tabular}{l|c|c|c|c|c}
\toprule
Approach &  Age of Topic & Popularity & Age of the Article  & Alphabetical & Geo-Location (TREC 21)  \\
\midrule
\hline
\multicolumn{6}{c}{SPLADE} \\
\hline

GEP &  \textbf{0.2624} & 0.2013 & \textbf{0.2129} & \textbf{0.2041} & \textbf{ 0.3543}\\
SCS   &  0.3738* & 0.2243 & 0.2782* & 0.2438* & 0.5261*\\
AvIDF &  0.3995* & 0.2128 & 0.2889* & 0.2471* & 0.582*\\
 AvICTF &  0.4088* & \textbf{0.1884} & 0.2714* & 0.2469* & 0.5695*\\
AvPMI &  0.3713*  & 0.2083 & 0.2758* & 0.2414* & 0.5474* \\
CORI  &  0.4062* &  0.1888 & 0.2735* & 0.2467* & 0.5631* \\

\hline
\multicolumn{6}{c}{BM25} \\
\hline

GEP &  \textbf{0.2880} & \textbf{0.1567} & \textbf{0.1759} & \textbf{0.1335} & \textbf{0.2449}*\\
SCS   &  0.3505* & 0.1620 & 0.2336* & 0.1407 & 0.4134*\\
AvIDF &  0.3726* & 0.1621 & 0.2443* & 0.1436 & 0.4844*\\
 AvICTF &  0.3773* & 0.1706* & 0.2305* & 0.1431 & 0.4668*\\
AvPMI &  0.3562*  & 0.1645 & 0.2302* & 0.1413 & 0.4388* \\
CORI  &  0.3735* &  0.1703* & 0.2338* & 0.1422 & 0.4582* \\

\hline
\multicolumn{6}{c}{TF-IDF } \\
\hline
GEP &  \textbf{0.2886} & \textbf{ 0.1583 } & \textbf{0.1759} & \textbf{0.1336} & \textbf{0.2454}*\\
SCS   &  0.3512* & 0.1628 & 0.2339* & 0.1420 & 0.4138*\\
AvIDF &  0.3734* & 0.1629 & 0.2447* & 0.1449  & 0.4847*\\
 AvICTF &  0.3780* & 0.1716* & 0.2307* & 0.1445 & 0.4671*\\
AvPMI &  0.3568*  & 0.1652 & 0.2304* & 0.1423&  0.4390* \\
CORI  &  0.3726* &  0.1708* & 0.2331* & 0.1435 & 0.4586* \\
\bottomrule

\end{tabular}
\end{adjustbox}
\vspace{-1\baselineskip}
    \label{tab:results_bm25}
\end{table*}


\begin{table*}[t]
\caption{Evaluation of the predictors over common first-pass retrieval rankers
with different Query Expansion models applied for rankings with k=100. * indicates significant differences (t-test, with Bonferroni correction, p$<$0.01) in prediction performance \iocrc{between} GEP \iocrc{and} the baselines. }
    \centering
   \begin{adjustbox}{width=0.99\textwidth}
  \begin{tabular}{l|c|c|c|c|c}
\toprule
Approach &  Age of Topic & Popularity & Age of the Article  & Alphabetical & Geo-Location (TREC 21)  \\
\hline
\multicolumn{6}{c}{BM25 + RM3 } \\
\hline
GEP &  \textbf{0.3134} &  \textbf{ 0.1722} & \textbf{0.2030} & \textbf{0.1456} & \textbf{0.2723}\\
SCS   &  0.3772* &0.1763 & 0.2507* & 0.1548 & 0.4364*\\
AvIDF &  0.3941* & 0.1746 & 0.2603* & 0.1571  & 0.5038*\\
 AvICTF &  0.3986* & 0.1835 & 0.2493* & 0.1567 & 0.4876*\\
AvPMI &  0.3823*  & 0.1747 &  0.2481* & 0.1561 &  0.4613* \\
CORI  &  0.3940* &  0.1836 & 0.2514* & 0.1557 & 0.4796* \\

\hline

\multicolumn{6}{c}{BM25 + KLQ } \\
\hline

GEP &  \textbf{0.2962} & \textbf{0.1625} & \textbf{0.1911} & \textbf{0.1430} & \textbf{0.2614}\\
SCS   &  0.3624* & 0.1706 & 0.2491*& 0.1489 & 0.4226*\\
AvIDF &  0.3813* & 0.1670 & 0.2598* & 0.1513  & 0.4915*\\
 AvICTF &  0.3864* & 0.1708 & 0.2467* & 0.1510 & 0.4750*\\
AvPMI &  0.3669*  &  0.1651&  0.2461* &  0.1505 &   0.4479* \\
CORI  &   0.3815* &  0.1708 &  0.2487* & 0.1500 & 0.4667* \\
\hline

\multicolumn{6}{c}{SPLADE + RM3 } \\
\hline

GEP &  \textbf{0.1912} & \textbf{0.2870} & \textbf{0.2549} & \textbf{0.1574} & \textbf{0.3511}\\
SCS   &  0.2365* & 0.3300* & 0.2871*& 0.1670& 0.5149*\\
AvIDF &  0.2438* &  0.4150* &  0.2898* & 0.1637& 0.5665*\\
 AvICTF &  0.2293* & 0.4241* & 0.2895* & 0.1576 & 0.5564*\\
AvPMI &  0.2344*  & 0.3857*&  0.2830* &  0.1610 &  0.5321* \\
CORI  &  0.2315* &  0.4216* &  0.2895* & 0.1572 & 0.5508* \\
\hline



\multicolumn{6}{c}{SPLADE + KLQ } \\
\hline

GEP &  0.2291 & \textbf{0.2410} & \textbf{0.2662} & \textbf{0.1777} & \textbf{0.3251}\\
SCS   &  0.2200 & 0.3300* & 0.2980*& 0.1738& 0.4961*\\
AvIDF &  0.2161 &  0.3538* &  0.3009* & 0.1847*& 0.5516*\\
 AvICTF &  \textbf{0.2120*} & 0.3613* & 0.3009* & 0.2054* & 0.5395*\\
AvPMI &  0.2230  &  0.3245*&  0.2933* &  0.1932* &   0.5163* \\
CORI  &  \textbf{0.2120}* &  0.3615* &  0.3010* & 0.2045* & 0.5335* \\
\hline



\bottomrule

\end{tabular}
\end{adjustbox}
\vspace{-1\baselineskip}
    \label{tab:results_PRF}
\end{table*}

\looseness -1 \noindent\textbf{Results for RQ1:} To answer RQ1, we evaluate the performance of our GEP predictor \ya{in terms of} its ability to predict the exposure that groups of documents \ya{will} receive in a ranking of size $k\text{=}100$.
We \ya{generate} rankings with different retrieval models namely BM25, TF-IDF and SPLADE. We calculate the actual exposure \iocrc{values} the documents receive and compare \iocrc{these} with the predictions made by our proposed approach and the baselines for each of the categories (where a category, such as \iocrc{Geo-Location}, is a set of related groups, such as Europe, Asia, etc.). 

Table~\ref{tab:results_bm25} shows the results of the evaluation when a first-pass retrieval model is used for \iocrc{ranking}. The table reports the Jensen-Shannon distance between the prediction and the actual exposure of each of the groups. The optimal prediction perfectly estimates the actual exposure and would result in a distance of 0. From the table, it \ya{is} apparent that our \iocrc{GEP} approach achieves the best results on all the groups when BM25 or TF-IDF is used as a ranker. GEP significantly outperforms (paired t-test, with Bonferroni correction, p$<$0.01) the baselines on three out of the five categories. The differences are significant for the Geographic Location, the Age of Topic, as well as the Age of Article \ya{categories}. Analysing the Popularity category, we note that our predictor is significantly \bj{better in predicting the exposure } compared to the  AvICTF and the CORI baselines. \cj{The biggest differences between the predictors can be observed in the geographic location category, on which our prediction outperforms the next best predictor by up to $\sim$40\%.} Geographic location is the category with the highest number of groups. When SPLADE is used as a retrieval model, we observe that our \bj{\iocrc{GEP} predictor} gives the best prediction on four out of five categories. In all of these four categories, our predictor significantly outperforms the baselines. GEP is only outperformed in the Popularity category. However, there is no \ya{observed} statistical significance. 
The\gm{se} results show that our approach \ya{is overall} the most effective predictor compared to all other evaluated \ya{baselines}. \ya{The} TF-IDF retrieval model \ya{\iocrc{also} leads to} the same trends as with BM25. 

To further investigate the behaviour of the predictors, we consider how dispersed the exposure is distributed over the groups in the actual rankings. We have identified the coefficient of variation (CV)~\cite{abdi2010coefficient}, which is the ratio of the standard deviation of a distribution $\sigma$ to the mean $\mu$ of the distribution to be a suitable measure for this, where: $C V=\frac{\sigma}{\mu}$\gm{.} We express the CV as a percentage. A lower CV indicates a flatter distribution, e.g., if the actual exposure is distributed [0.25, 0.25, 0.25, 0.25] over four groups \gm{then $CV=0$}. 

\begin{figure*}
  \centering
    \includegraphics[width=0.7\textwidth]{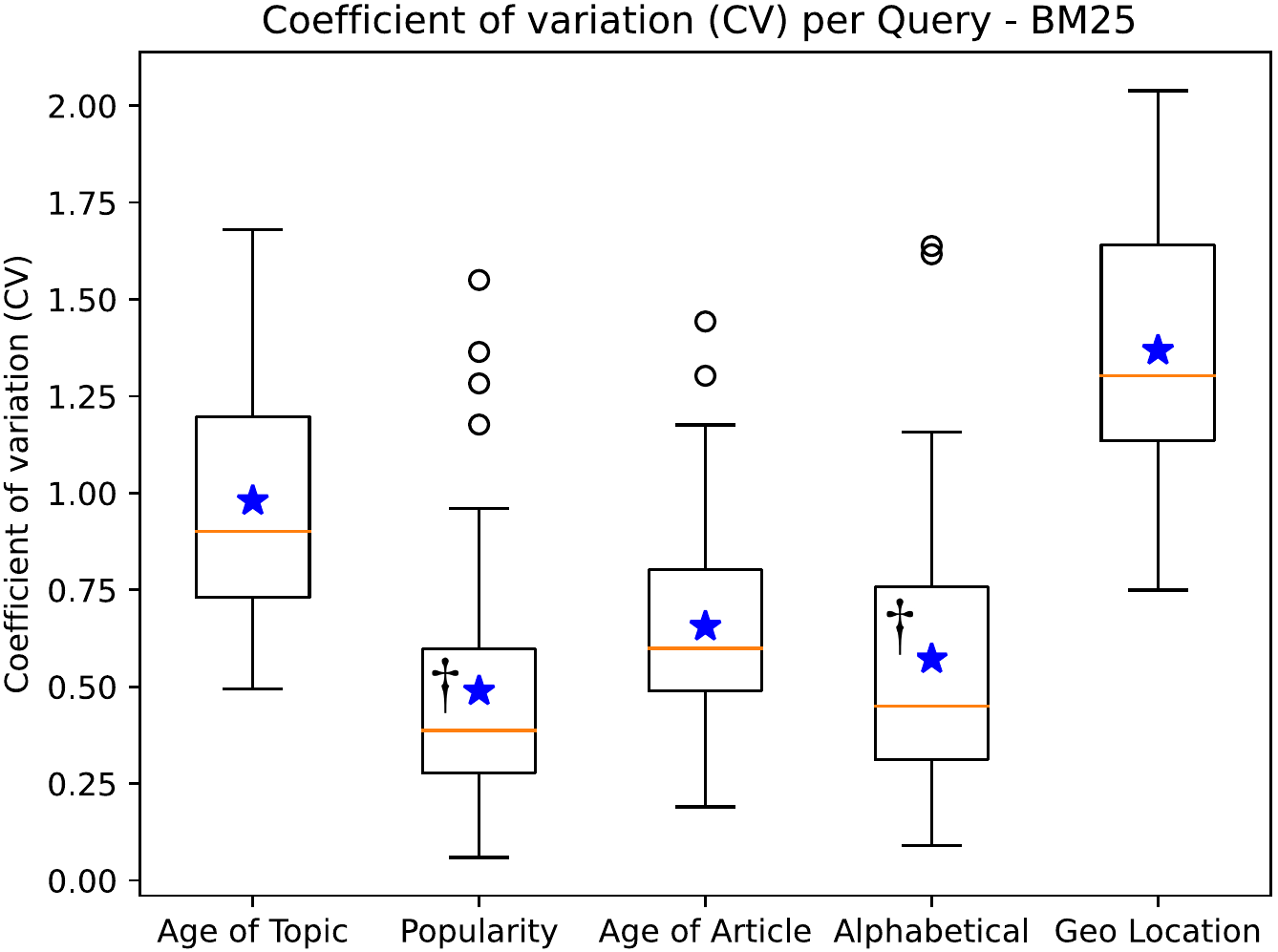}
    \caption{Dispersion of the exposure in the categories per query measured by Coefficient of Variation. We use $\dagger$ to indicate significant differences (t-test, p$<$0.05) to the Age of Article category values. }
     \label{fig:boxplots}
\end{figure*}

Figure~\ref{fig:boxplots} shows the distribution of \iocrc{the} CV values over all queries per category. Analysing the figure, it \ya{is clear} that the categories Popularity and Alphabetical have the lowest median (orange line) and mean (blue star) CV. The category Geographic Location has the highest median and its interquartile range contains the highest CV values. Notably, all categories on which we significantly outperform the baselines show a higher variability in the distribution of the actual exposure scores in the ranking than the other categories, \bj{which suggests that our predictor is better than the baselines when the exposure is not uniformly distributed}. We use statistical testing to \iocrc{ascertain} whether the differences between the distributions of the two categories with the lowest CV values and the other categories are significant (t-test, p$<$0.05). We compare the category Age of Article (which has the lowest CV mean of the groups on which our predictor outperforms the baselines) against the Popularity and Alphabetical categories (for which we \iocrc{observe} visible but non-significant improvements \ya{in} Table~\ref{tab:results_bm25}). The results of our statistical tests show a significant difference, denoted as $\dagger$, between Age of Article and the two other categories in terms of the distributions of CV-values. This \ya{might indicate} that the coefficient of variation plays an important role in predicting the exposure over groups since all of our baselines \ya{provide} better estimates for categories with more flat distributions. We leave \ya{the} investigation of this interesting \ya{finding} to future work. To answer RQ1, we conclude that our proposed predictor can outperform all of \ya{the used} baselines and provide \ya{promising} results independent of the deviations in the exposure distribution.  

\vspace{-1\baselineskip}


\tj{
\noindent\textbf{Results RQ2:} To answer RQ2, we investigate how the prediction performance is affected when a query is altered through Pseudo-Relevance Feedback. \cj{The first two sections of Table~\ref{tab:results_PRF} show the results for the predictors when BM25 is combined with different PRF strategies (RM3, KLQ).}  We observe that for both PRF \ya{approaches} over all categories, our predictor \bj{GEP} outperforms the baselines. On three categories, our approach provides significant improvements \bj{over the baselines}\gm{.} This is the same trend we have \ya{observed} when analysing the standard retrieval approaches. However, the results \iocrc{in Tables} \ref{tab:results_bm25} and \ref{tab:results_PRF} show that for all of the categories and approaches, the prediction quality drops compared to the standard retrieval rankings. This is to be expected since the predictors have no knowledge of how the query might be changed. Nevertheless, the drop \iocrc{in} \bj{performances} is marginal\gm{,} indicating that the effect of PRF on the exposure is very limited and that our predictor is still useful even when the query is altered. When SPLADE is combined with RM3, our predictor outperforms the baselines on all categories with significant differences on four out of five categories. 
Analysing the ranking with \iocrc{the} KLQ expansion, we observe statistical differences to all baselines on three out of five categories, with GEP providing the lowest numbers. On the age of topic category, the CORI and the  AvICTF baseline produce the best prediction, \bj{outperforming} our approach. On the Alphabetical category, however, GEP achieves the best scores. Overall, we observe that our predictor remains robust with only small drops in the prediction quality even when PRF is applied in the pipeline. A possible explanation for the robustness of our predictor when PRF is applied and unknown query terms are added can be found by revisiting Figures~\ref{fig:possibleExpPerm}. First \ya{let us} assume a scenario where the PRF component leads to only a re-ranking of documents, and does not add any new documents to the ranking. \bj{Assuming such a ranking, we can observe from Figure~\ref{fig:possibleExpPerm} how the exposure can change per group by looking at the individual bell curves. If the number of documents in a group stays the same, an individual bell curve shows \iocrc{the} range of how much exposure a group can \iocrc{obtain}.} From the figure, it is \ya{clear} that the range of possible exposure values is very restricted and narrows down even more for an increasing number of documents. Explanations for a second scenario where the exposure changes with variations in the number of documents per group due to \iocrc{alterations} of the query can also be found in  Figure~\ref{fig:possibleExpPerm}. The figure shows that substantial changes in the exposure distribution only occur if a relatively large number of documents (e.g., +5) from a group are added to the ranking. Moreover, as can be seen from Figure~\ref{fig:exposureCombined}(a), if the documents are added in lower \iocrc{ranking} positions, then relatively little additional exposure is accumulated, as previously discussed in Section 3.
To answer RQ2, we can conclude that even when PRF is applied, our proposed predictor is still effective.  Moreover, we have outlined two possible explanations, \iocrc{for} why PRF interventions might only \ya{lead to} limited effects on the exposure \iocrc{among} groups. \bj{We have \iocrc{also} conducted \iocrc{additional PRF} experiments using the Bo1 model from the Divergence from Randomness framework~\cite{amati2002probabilistic}, and  axiomatic query expansion (AEQ)~\cite{fang2006semantic} and have observed that all the observations from RQ2 also hold for \iocrc{these}. Due to page limitations, we \iocrc{do not} present these results in this \iocrc{paper}.
}}

\section{Conclusions}\label{sec:conclusion}

In this \iocrc{paper}, we have introduced \io{the} novel task of \io{query exposure prediction}. We have conducted a detailed investigation of the exposure allocation problem and have proposed a \iocrc{new} predictor, GEP, which has been explicitly designed for query exposure prediction. \tj{Our extensive experiments on two standard test collections of the TREC Fair Ranking Track showed that our predictor is a suitable \io{and effective} choice for the task of query exposure prediction.} GEP outperforms the \io{used} baselines across multiple retrieval models, even when the knowledge of the query is limited, i.e., when PRF is applied. Our evaluation shows that our predictor achieves improvements of up to $\sim$40\% in prediction accuracy over the next best performing baseline. The main aim of this work is to present \io{this important} task to the community and to encourage the development of \iocrc{further} exposure predictors. \fj{In future work, we aim to develop end-to-end predictors across the entire pipeline,  exploring potential avenues such as the development of a supervised learning approach trained on exemplary rankings.}
%
%
%
\bibliographystyle{splncs04}

\end{document}